\documentstyle[12pt]{article}
\def\be{\begin{equation}}
\def\ee{\end{equation}}
\def\({\left (}
\def\){\right )}
\def\phih{\hat{\varphi}}
\def\phid{\varphi^{\dagger}}
\def\phihd{\phih^{\dagger}}
\def\tg{\tilde{g}}\def\tl{\tilde{\lambda}}

\begin{document}
 \begin{titlepage}
 \vspace*{-4ex}
 \null \hfill LBL--33067\\
\null \hfill October, 1992\\
\vskip 2 true cm
 \begin{center}
 {\bf \LARGE On the High $T$ Phase Transition in the
Gauged $SU(2)$ Higgs Model}\\[11ex]
{\large
  Vidyut Jain and Aris Papadopoulos}
   \\ [5ex]
%

{\large \it
Theoretical Physics Group\\ Lawrence Berkeley Lab\\[0.1cm]
   Berkeley CA 94720}\\ [2ex]
 \end{center}
\vskip 2.5cm
 \begin{abstract}
 We study the effective field theory of a weakly coupled
3+1d gauged $\phi^4$ type
 model at high temperature. Our model has $4N$ real scalars
 ($N$ complex Higgs doublets) and a gauge group $SU(2)$ which
is spontaneously broken by a nonzero scalar field $vev$ at zero
temperature.
 We find, for sufficiently large $N$, that the
 transition from the high temperature
symmetry restoring phase to the
 low temperature phase can be either first order or
 second order depending on the ratio of the gauge coupling
 to the scalar self coupling.
 \end{abstract}
 \end{titlepage}

Recently there has been much interest in the nature of the
high temperature electroweak phase transition. It has been known
for some time that at high enough temperature the ground state of
the electroweak model is symmetry preserving \cite{DJ}, even if the zero
temperature ground state is not. What is not well understood is
how the transition from the high temperature phase to the low
temperature
phase proceeds.
If the phase transition is first order, it may be
possible to generate the baryon asymmetry of the universe (BAU) within
the electroweak model itself \cite{Del},\cite{Sel},\cite{Shop},

It is currently believed that the phase transition is first order,
but too
weakly first order for the purposes of generating the
BAU \cite{ls},\cite{BB}. A one--loop
high temperature calculation suggests that the electroweak phase
transition
is first order \cite{Greg}.
However, the high temperature field theory suffers
from infrared divergence problems. A partial resummation of higher loop
contributions implies that the strength of the first order transition
is reduced relative to the one--loop result \cite{car},\cite{ls},
\cite{BB},\cite{pa}.

In this letter we would like to point out that higher loop
contributions
may not only reduce the strength of the first order transition but
actually result in a second order phase transition. We study not the
full electroweak theory but a weakly coupled gauged $\phi^4$ model.
The gauge group is $SU(2)$ and we work with $4N$ real scalars
($N$ complex Higgs doublets). Each of the $SU(2)$ Higgs doublets
is coupled to the $same$ gauge fields.

Our analysis follows \cite{vj}.
The 3+1d model is, at high temperature, formally
equivalent to a euclidean field theory with one compact dimension. For
physics at scales less than $O(T)$ it is sufficient to study the
effective 3d Lagrangian that results from integrating out the compact
dimension. Vacuum polarization effects can be computed exactly
in three dimensions, in contrast to four dimensions.
Furthermore, the explicit $\phi^4$ term can be removed by
introducing a dimension two auxiliary field $\chi$ \cite{Coleman},
\cite{root},\cite{ml}. The auxiliary
field allows for a straightforward computation of all the quantum
corrections that survive in the limit of arbitrarily large $N$. The
corrections that survive depend on how the gauge coupling and scalar
self coupling are held fixed as $N$ increases.

We find that if $g^2/\lambda\leq O(1)$, where $g$ is the 4d gauge
coupling and $\lambda$ is the scalar self--coupling, the effective
scalar
potential $V_{eff}$ admits only a 2nd order phase transition at
large $N$.
This is because vacuum polarization effects at nonzero external
momentum
prevent $any$ $T\phi^3$ term from appearing in $V_{eff}$. This is
in contrast to the case $g^2/\lambda\sim O(N)$ for which a first
order phase transition is still possible.

These conclusions are
the same as for the high temperature abelian Higgs model studied
by one of the current authors in \cite{vj}. The results are the same
because at large $N$ and weak coupling
the nonabelian nature of the $SU(2)$ group
is not important and the three $SU(2)$ gauge fields interact with the
scalars like three abelian gauge fields.

In the abelian case, it was shown that the next--to--leading
corrections compete with the leading corrections in the $1/N$
expansion when $N=1$ and $g^2/\lambda\sim 1$. Therefore, one
may question the applicability of the our results here to the
physically
interesting situation of one Higgs doublet, especially since the
number
of $SU(2)$ gauge fields is just one less than the number of scalars
in this case.
However, the leading order
and next--to--leading order result for $V_{eff}$ just corresponds to
resuming some infinite classes of Feynman
diagrams in a way that avoids overcounting any of them. For example,
for
$g^2/\lambda\leq O(1)$,
the $O(1)$ result for $V_{eff}$ that we derive below includes the
diagrams of Figure 1. The fact that in a systematic approach the phase
transition is second order at $O(1)$ in $V_{eff}$ when $g^2/\lambda\leq
O(1)$ means that any first order behaviour comes from at best
$O(1/N)$ terms in $V_{eff}$. Therefore, all $O(1/N)$ terms must be
computed in a systematic way in order to make a reliable statement
about the phase transition in the $N=1$ case. Such a computation
does not currently exist in the literature, even for the abelian
Higgs model. We are currently trying to develop a different $1/N$
expansion that we feel will be more suited to the electroweak model.
Namely, to work with $N$ real scalars and the four $SU(2)\times U(1)$
gauge fields in $N$ dimensions. For $N$ bigger than four, such a
model is nonrenormalizable but can be treated as an effective low
energy theory.

What
we present below is a straightforward generalization of the results of
\cite{vj}. For
technical details, the reader is referred to \cite{vj}.
  Related works, in the case of a pure scalar theory, are
references \cite{root},\cite{vjb}, \cite{fabio} and \cite{tetra}.
The 3d effective
field theory of high temperature  gauged $\phi^4$ models has been
previously studied using the $\epsilon$--expansion technique in ref.
\cite{gp}, and more recently in \cite{mr}.
Recently, the abelian Higgs model has also been studied
in \cite{buch} using a different method than the one considered in
\cite{vj}. The authors of \cite{buch} establish a range in
parameter space $(\lambda,g)$ where the phase transition is first order,
although no definite prediction of a second order transition is
made for parameter values outside this range.

\newpage
The tree level Lagrangian we consider here is
\def\A{{\bf A}}
\def\F{{\bf F}}
\be L [\phi,\A] := N D_\mu\phi^{\dag}_A D^\mu \phi^A -
{1\over 2}{\rm Tr}\;
     \F_{\mu\nu}\F^{\mu\nu} - V(\phi), \label{e1} \ee
where the tree potential is
\be V(\phi) := {N\lambda\over 4} \(\phi^{\dag}_A\phi^A-v^2\)^2 ,\ee
and the kinetic terms can be expanded using
\be D_\mu\phi_A := \(\partial_\mu+ {ig\over \sqrt{N}}\A_\mu\)\phi_A \ee
and
\be \F_{\mu\nu} := \partial_\mu\A_\nu-\partial_\nu\A_\mu
        + {ig\over\sqrt{N}}[\A_\mu,\A_\nu]. \ee
Our conventions are as follows. This is the 3+1d Minkowski space
Lagrangian with metric $\eta_{\mu\nu}:=$diag$(+,-,-,-)$.
The indices $A,B$, etc., run from
$1$ to $N$. Each field $\phi^A$ is a complex Higgs doublet. The
gauge and scalar couplings are $g$ and $\lambda$, respectively.
The three $SU(2)$ gauge fields $A^a_\mu$, $a=1,2,3,$  are defined via
$\A_\mu:=A^a_\mu T_a,$ where $T_i$ are half the pauli matrices.
The purely scalar part of the Lagrangian is normalized to have
an overall factor of $N$ only for later convenience; the kinetic
term can be put in a canonical form by the rescaling $\phi\rightarrow
\phi/\sqrt{N}$. Finally, we remark that
in order to avoid any problems associated with triviality \cite{WB} we
treat
our model as an effective low energy theory valid below some scale
$\bar{\Lambda}$.

To study the high temperature phase transition of the model
(\ref{e1}) we follow the procedure outlined in \cite{vj}.
We first
write the corresponding euclidean version of (\ref{e1}). The
euclidean
coordinates are $(\tau,\vec{x})$, where at high temperature $\tau$
describes the compact dimension. All fields are periodic in $\tau$
with period $\beta=1/T$. At sufficiently high $T$ the model is
effectively three dimensional, a fact that we exploit below.

The four dimensional fields can be
expanded as
\be \phi^A(\vec{x},\tau) =\sum_n \phi^A_n(\vec{x}) \psi^n(\tau),
   \qquad  \A^\mu(\vec{x},\tau)=\sum_n \A^\mu_n(\vec{x})\psi^n(\tau)
\ee
where $\psi^n$ are a complete set of periodic functions on the circle,
$\psi^n(0)=\psi^n(\beta)$. Thus, each 4d field yields an infinite
tower of 3d fields when the compact dimension is integrated out
in the action.
For what we are
interested in only the zero modes $n=0$, for which $\psi^0(\tau)=
\psi^0(0)$, are important in the effective 3d model. This is because
for $n\neq 0$ the fields $\phi^A_n, \A^\mu_n$ pick up large nonzero
masses
of $O(T)$ from the 4d kinetic term. Truncating the
spectrum to keep only the zero modes and integrating out the
compact dimension gives the effective 3d Lagrangian
\def\cA{{\cal A}}
\def\cF{{\cal F}}
\def\brho{\rho\hspace{-0.19cm}\rho}
\def\bvth{\vartheta\hspace{-0.225cm}\vartheta}
\begin{eqnarray}
L[\varphi,\cA,\rho]&:=& -N \phid \vec{\partial}^2\varphi
+{N\tilde{\lambda}\over 4}(\phid\varphi-{\tilde{v}}^2)^2 \nonumber \\
 & & + {\rm tr}\left((D_m\brho)^2+{1\over2} {\cF}_{mn}{\cF}^{mn}\right)
 +\tilde{g}^2\phid(\cA_m\cA^m+\brho^2)\varphi\nonumber \\
& &
+i\tilde{g} \sqrt{N}
\left(\partial_m \phid \cA^m\varphi-\phid\cA^m\partial_m\varphi\right)
\end{eqnarray}
where we have defined the three dimensional fields and coupling
constants
\begin{eqnarray}
&\varphi^A:=\sqrt{\beta}\phi^A_0,\;\;\;
\tilde{\lambda}:=\lambda/\beta,\;\;\;{\tilde{v}}^2:=\beta v^2\nonumber\\
&\cA^m:=\sqrt{\beta}\A^m_0,\;\;\;\brho:=\sqrt{\beta}\A_0^0,\;\;\;
\tilde{g}:=
g/\sqrt{\beta}
\end{eqnarray}
in terms of the zero modes of the four dimensional euclidean fields
and the four dimensional coupling constants. We have suppressed indices
labeling the number of Higgs doublets so that, for example,
$\phid\varphi
=\phid_A\varphi^A$. The indices $m,n$ are used for the euclidean
spatial coordinates. The 3d field strength is ${\cal F}_{mn}
=\partial_m\cA_n-\partial_m\cA_m+ig[\cA_m,\cA_n]/\sqrt{N}$. The
3d gauge covariant derivative on the scalars $\brho$ is
$D_m\brho=\partial_m\brho+i\tg[\cA_m,\brho]/\sqrt{N}$.
Finally, $\cA_m=\cA^m,\partial_m=\partial^m$.

Therefore, the three dimensional model we study is a gauged Higgs
model with an extra
triplet of real scalars $\brho:=\rho^a T_a$ in the adjoint
representation
of $SU(2)$. These extra scalars are just the zero modes of the
4d longitudinal gauge bosons, while the 3d gauge bosons are just the
zero modes of the 4d transverse gauge bosons.

Before proceeding we would like to stress some important points.
As shown in \cite{gp}, if one integrates out the nonzero modes
$n\neq 0$
at the quantum level rather than just truncating the spectrum, there
is a finite $O(T)$ correction to the mass of $\varphi$. This
correction is very important, indeed it is the term that gives
symmetry restoration at high enough temperature. We will obtain
this term another way, in analogy with what one does in effective
low energy theories of the strongly interacting standard model
or effective four dimensional supergravity models inspired by
string theory \cite{mk}.

The three dimensional field theory will be
divergent and it must be regulated.
In fact, the only divergent integrals that will
be important are linearly divergent and we will regulate them by
simply using  sharp momentum cutoffs.
We then give a physical interpretation
to the cutoffs, i.e. the scale at which the full four dimensional
physics becomes important. This scale is of $O(T)$.
This interpretation is in complete analogy with, for example,
the effective
4d theories where the regulating scale is taken to be of order
the compactification scale. In our case the identification
of the cutoffs with $T$ can be
made precise because the corresponding 4d corrections are well
known.

In \cite{vj} it was shown for the pure scalar case that if one
regulates all divergent integrals with the same cutoff $\Lambda$ then
the identification $\Lambda=\pi^2 T/6$
reproduces exactly the $T^2$ results from the four dimensional
integrals. In general, for a model with different types of particles
and different couplings, one should use different cutoffs for the
different divergent terms that appear in $V_{eff}$ \cite{mk}. This
is because the 3d effective theory does not know the precise way
in which the full four dimensional physics enters at scales of $O(T)$.
In \cite{vj} it was also shown that, in order to obtain the correct
temperature dependent electric mass for the 4d abelian gauge field, a
different cutoff $\Lambda'=\pi^2 T/3$ must be used to regulate the
linearly divergent vacuum polarization of the zero mode of the 4d
longitudinal gauge field. All of this will be seen more explicitly
below.

We remark that since the four dimensional model (1) is
only valid up to a scale $\bar{\Lambda}$ we must require
$\Lambda,\Lambda'<\bar{\Lambda}$, i.e. that $T$ is sufficiently small.
Finally note that the 4d effective potential $V_{eff}(\phi)$
can be obtained
from the three dimensional potential $V_{eff}(\varphi)$
by dividing by $\beta$ and using (7).

To calculate the dominant corrections to $V_{eff}$ it is now
convenient to
introduce a dimension two auxiliary (real) field $\chi$
\cite{Coleman},
\cite{root}, \cite{ml}. This technical
trick allows one to write a closed expression for the corrections to
$V_{eff}$ that do not fall with $N$. Specifically, we work with the
3d Lagrangian
\begin{eqnarray}
   L[\varphi,\chi, \cA,\rho] &:=& L[\varphi,\cA,\rho]
      -{N\over \tilde{\lambda}}
     \left(\chi-{\tilde{\lambda}\over 2}(\phid\varphi-\tilde{v}^2)
\right)^2 ,
    \label{e8}
 \end{eqnarray}
in which the potential piece becomes
\be
V(\varphi, \chi)
  =-N{\chi^2\over\tilde{\lambda}}+N\chi(\phid\varphi-\tilde{v}^2).
\ee
The original form (3) can be recovered by using the equation of motion
for $\chi$.

To calculate $V_{eff}(\varphi)$ we proceed as follows
\cite{jw},\cite{Coleman}. First, we expand (\ref{e8}) about spatially
constant backgrounds $\varphi^A,\chi$ thus
\def\hvp{\hat{\varphi}}
\be \varphi^A\rightarrow\varphi^A +{\hvp^A\over\sqrt{N}}, \qquad
      \chi\rightarrow\chi+{\hat{\chi}\over\sqrt{N}}, \ee
deleting all terms linear in the quantum fields $\hvp,\hat{\chi},
\cA$ and $\rho$. We need not keep backgrounds for $\cA$ and $\rho$
because we are only interested in the effective scalar potential.
To this we must add gauge fixing and ghost terms. We chose
\be
 L_{g.f.} = {1\over 2\alpha}
         \left(\partial^m \cA_m^a+i\tilde{g}\alpha[\phihd T^a\varphi
               -\phid T^a\phih]\right)^2
\ee
and
\begin{eqnarray}
 L_{ghost}& =&\partial^m\bar{\vartheta}^aD_m\vartheta_a
              +{1\over2}\tilde{g}^2\alpha \bar{\vartheta}^a
              \phid \varphi \vartheta_a
            \nonumber \\
          &&+{\tilde{g}^2\alpha\over\sqrt{N}}
             (\phihd \bvth\bvth^{\dag} \varphi
              +\phid \bvth^{\dag}\bvth \phih)
\end{eqnarray}
where $\alpha$ is an arbitrary (small) parameter and
$\vartheta^a$ are
the grassmanian ghost fields ($\bvth=\vartheta^a T_a$). The gauge
index $a$
runs from 1 to 3.
Also $D_m\bvth=\partial_m\bvth+i\tg[\cA_m,\bvth]/\sqrt{N}$.

Altogether, this procedure defines
a quantum Lagrangian given by
\begin{eqnarray}
 L[\hvp,\hat{\chi},\cA,\rho,\vartheta] &:=& -{N\chi^2 \over\tl}
+N\chi(\phid
 \varphi-{\tilde{v}}^2)
 +\phid\phih\hat{\chi}+\phihd\varphi\hat{\chi}
 +\frac{1}{\sqrt{N}}\phihd\phih\hat{\chi}\nonumber \\
& &+\phihd(-\vec{\partial}^2+\chi)\phih- {N\hat{\chi}^2\over\tl}
     \nonumber \\
& &+{\rm tr}\left((D_m\brho)^2+{1\over2}{\cF}_{mn}{\cF}^{mn}\right)
 +\tilde{g}^2\phid(\cA_m\cA^m+\brho^2)\varphi\nonumber \\
& &+{1\over2\alpha}(\partial^m\cA_m^a)^2
 -{1\over2}\alpha\tilde{g}^2(\phihd T^a\varphi-\phid T^a \phih)^2
\nonumber \\
 & &+{\tilde{g}^2\over\sqrt{N}}[\phihd\cA^m\cA_m\varphi+\phid\cA^m
\cA_m\phih]
    +{\tilde{g}^2\over N}\phihd(\cA^m\cA_m+\brho^2)\phih
\nonumber \\
& &+{i\tilde{g}\over \sqrt{N}}
\left(\partial_m \phihd \cA^m\phih-\phihd\cA^m\partial_m\phih\right)
+L_{ghost}
\label{e12}
\end{eqnarray}

We then calculate $V_{eff}(\varphi,\chi)$ as a sum of the tree--level
potential plus all one--particle--irreducible (1PI) diagrams of
(\ref{e12}). The effective scalar potential $V_{eff}(\varphi)$
is found by eliminating $\chi$ through its equation of motion,
$\partial V_{eff}(\varphi,\chi)/\partial\chi=0$, which gives $\chi$
as a function of $\varphi$. Since $V_{eff}$ is the sum of 1PI
diagrams it need not be convex \cite{slade},\cite{alex}. $V_{eff}$ will
allow us to deduce if the phase transition is 1st or 2nd order
(see also \cite{zj}).

As mentioned, the contributions to $V_{eff}$ that survive as
$N\rightarrow
\infty$ depend on how $\tl$ and $\tg$ are held fixed in this limit.
The two interesting cases that we consider here are
\begin{eqnarray}   i) & & \; \; \tl\; \;,\quad\tg\quad {\rm fixed}
\nonumber \\
    ii) & & (\tl N),\;\tg\quad {\rm fixed}. \nonumber
\end{eqnarray}

Let us consider case $i)$ first. Then it is known \cite{DJ} that there
is a contribution of $O(N)$ to $V_{eff}$ from the purely scalar part
of the model which comes from summing the ``superdaisy'' graphs
in ordinary perturbation theory in $\hbar$. There are no contributions
of $O(N)$ involving the gauge fields \cite{vj}. The gauge fields first
contribute at $O(1)$ in $V_{eff}$ because the number of gauge degrees
of freedom does not increase with $N$. From examining (\ref{e12}) it
is clear that the $O(N)$ quantum correction comes from the part which
is quadratic in $\hvp$ --- the quantum fields $\hat{\chi},\cA$ and
$\rho$ can be ignored. A one--loop computation gives $(\chi\geq 0)$
\begin{eqnarray}
    V_N(\varphi,\chi) &=& -N {\chi^2\over \tl}+N\chi(\phid\varphi-
     \tilde{v}^2)+2N{\rm Tr}\ln(-\vec{\partial}^2+\chi)
 \nonumber \\
    &=& -N{\chi^2\over\tl}+N\chi(\phid\varphi-\tilde{v}^2) +
          N{\Lambda\chi\over \pi^2}-N{\chi^{3\over2}\over 3\pi},\
\end{eqnarray}
where we have used a cutoff $\Lambda$ and dropped constants.
The identification $\Lambda = \pi^2 T/6$ reproduces exactly the
renormalized high temperature 4d results of \cite{DJ} for the
effective mass at the origin. The ``mass--gap'' equation for $\chi$
is then
\be  \chi={\tl\over2} \( \varphi^{\dag}\varphi-\tilde{v}^2
   +{T\over 6} -{\sqrt{\chi}\over 2\pi} \). \ee
The physical solution for $\sqrt{\chi}$
which then follows from $V_N$ is \cite{Coleman},\cite{root},\cite{vj},
\cite{vjb}
\be \sqrt{\chi} = {\tl\over 8\pi} \left [
\sqrt{1+{32\pi^2\over \tl}(\varphi^{\dag}\varphi-\tilde{v}^2+{T\over6}
        )} - 1\right]. \ee

The next--to--leading corrections to $V_{eff}$ involve all the
quantum fields. However, to $O(1)$, many of the terms in (13)
are unimportant. First note that $\phih$ appears in the quantum
Lagrangian quadratically. Thus these fields can be explicitly
integrated out. In order to do this we must first shift $\phih$
in such a way as to eliminate all the terms which are linear in
$\phih$. The effect of doing this to the order we are working in is
to simply replace the mixing term \cite{vjb}
\def\pd{\partial}
\be \phid\phih\hat{\chi}+\phihd\varphi\hat{\chi}\rightarrow
-\hat{\chi} {\phid\varphi\over (-\vec{\partial}^2+\chi)}\hat{\chi}. \ee
To $O(1)$,
no new contributions are generated from the $O(\tilde{g}^2\phih)$
mixing terms. Thus we can ignore these terms.

Integrating out the scalars generates $V_N$ and, to $O(1)$, a
gauge dependent contribution from the second term in the
fourth line in (13). It also generates, to  $O(1)$,
terms quadratic in the $\hat{\chi}$, $\brho$, and gauge fields, as
well as cubic and higher powers in these fields which appear
at higher order in the ${1\over N}$ expansion. The
quadratic contributions can be computed in terms of the
vacuum polarization
terms
$\Pi_{\hat{\chi}\hat{\chi}},$ $\Pi_{\rho\rho}$ and $\Pi_{mn}^{ab}$.
In calculating these  using the quantum Lagrangian
of eq.(13) we need to consider diagrams which only involve
$\phih$ fields in the loop.
Explicitly we have
\def\hc{\hat{\chi}}
\def\old{ \begin{picture}(77,22)(0,+8.5)
     \thicklines
  \put(4,10){\line(1,0){22}}\put(4,12){\line(1,0){22}}
  \put(36,11){\circle{20}}
  \put(46,10){\line(1,0){23}}\put(46,12){\line(1,0){23}}
\end{picture} }
\def\A{{\cal A}}\def\tg{\tilde{g}}\def\vt{\vartheta}
\def\ola{ \begin{picture}(71,28)(0,+2.5)
     \thicklines
  \multiput(4,11)(8,0){3}{\oval(4,4)[b]}
  \multiput(8,11)(8,0){3}{\oval(4,4)[t]}
\put(36,11){\circle{20}}
  \multiput(48,11)(8,0){3}{\oval(4,4)[b]}
  \multiput(52,11)(8,0){3}{\oval(4,4)[t]}
\end{picture} }
\def\olb{ \begin{picture}(59,28)(0,+2.5)
     \thicklines
  \multiput(4, 5)(8,0){7}{\oval(4,4)[b]}
  \multiput(8, 6)(8,0){7}{\oval(4,4)[t]}
\put(32,18){\circle{20}}
\end{picture} }
\def\olc{ \begin{picture}(77,22)(0,+8.5)
     \thicklines
  \multiput(4,6)(8,0){8}{\line(1,0){6}}
\put(36,16){\circle{20}}
\end{picture} }
\def\[{\left [}\def\]{\right ]}
\begin{eqnarray}
  \Pi_{\hc\hc}(\vec{k}) & = & \old =
\tl \int_{\vec{p}}
   {1\over [ \vec{p}^2+\chi] [  (\vec{p}+\vec{k})^2+\chi]  },
\nonumber \\
  \Pi_{\rho\rho}^{ab}(\vec{k}) &=& \olc = \tg^2\delta^{ab}
     \int_{\vec{p}}{1\over \vec{p}^2+\chi} ,
 \nonumber \\
  \Pi^{ab}_{mn}(\vec{k}) &=& \ola + \olb \nonumber \\
 &  & = \tg^2\delta^{ab} \int_{\vec{p}}\[
  {\delta_{mn}\over \vec{p}^2+\chi} -
     {{1\over2} (2p+k)_m (2p+k)_n
         \over[\vec{p}^2+\chi][(\vec{p}+\vec{k})^2
        +\chi]}\] .
  \label{vpe}
\end{eqnarray}
We recall that $m,n$ are spatial indices and $a,b$ are gauge indices.

It can be explicitly checked that all terms cubic or higher in the
remaining quantum fields do not generate any Feynman diagrams which
contribute at $O(1)$ to $V_{eff}$. This was also true for the abelian
Higgs model \cite{vj}, as well as for the purely scalar model
\cite{root},\cite{vjb}. The quadratic part of the action for the
remaining fields is
\def\ns{\hspace{-1.5cm}}
\begin{eqnarray}
    L[\hat{\chi},\cA,\brho,\bvth] & = &
     {1\over2} \rho_a\left[ (-\vec{\pd}^2+{1\over2}\tg^2\phid\varphi)
     \delta^{ab}+\Pi^{ab}_{\rho\rho}\right] \rho_b \nonumber \\
   & & \ns
   +  {1\over2} \cA_a^m\left[ (-\vec{\pd}^2+{1\over2}\tg^2\phid\varphi)
 \delta^{ab}\delta_{mn} +(1-\alpha^{-1})\partial_m\partial_n\delta^{ab}
       +\Pi^{ab}_{mn}\right] \cA_b^n \nonumber \\ & & \ns +
     \bar{\vartheta}^a [ -\vec{\pd}^2 +
            {1\over 2}\tg^2\phid\varphi]\vartheta_a
     \nonumber \\ & & \ns
     - \hat{\chi} \left [ \tl^{-1} + {\phid\varphi\over (-\vec{\pd}^2
        +\chi)} +\tl^{-1}\Pi_{\hc\hc} \right] \hat{\chi}.
\end{eqnarray}
$V_{eff}$ to $O(1)$ is then given by $V_N$
piece, a simple one--loop contribution from ghosts $\bvth$, and
the contributions from quadratic integrals over ${\cal A}$, $\brho$,
$\hat{\chi}$ with propagators modified by the vacuum polarization
effects (18). The $O(1)$ 4d gauge contributions correspond to summing
the 1PI diagrams given in Figure 1.

It is known [1],\cite{vjb}  that at order $N$,
the potential above  admits no 1st order phase transition. The $O(N)$
potential is exact in the small $\chi$ limit (up to 4d corrections).
To get the high $T$ 4d potential $V_{eff}(\phi)=T V_{eff}(\varphi)$ we
use (7). In the following $\phi^2:=\phi^{\dag}_A\phi^A$.

To $O(N)$, $dV_{eff}/d\phi^2 = \partial V_{eff}/d\phi^2= N\chi$. At
$\phi^2=0$ this vanishes at $T^2_2=6v^2$. For $T>\sqrt{6}v$
the origin is a global minimum, and at $T=T_2$ the potential grows
away from the origin. For $T>T_2$ the point $\chi=0$ is away
from the origin and this has the interpretation \cite{Coleman}
as the symmetry breaking minimum below $T_2$.

We now argue that, to $O(1)$, the phase transition is again second order.
Note that
all the expressions in (19) are diagonal in the gauge indices. Thus,
to $O(1)$, the nonabelian nature of the gauge fields is not felt; the
three $SU(2)$ gauge fields behave like three abelian gauge fields.
In \cite{vj} it was shown that the phase transition in the abelian
Higgs model is second order at $O(1)$ when $g^2/\lambda$ is held
fixed in the large $N$ limit. The reason for this is the following.

The case $g=0$ is known to have a second order transition to $O(1)$
\cite{vjb},\cite{vj}. Therefore, we need only examine the gauge
contributions.
At one--loop (in $\hbar$) and high temperature, each 4d gauge degree
of freedom gives a contribution $\sim - T \phi^3 $
to $V_{eff}$.
It was subsequently realised [6],[7],[9] that vacuum polarization
effects generate a large $T$ dependent mass for the longitudinal
gauge fields which eliminates their contribution to the cubic terms
in $V_{eff}$. This corresponds to the fact that the vacuum polarizations
for the zero modes of the 4d longitudinal gauge fields in (18) are
linearly divergent. However, it was also noted that vacuum polarization
effects do not generate such large $T$ dependent masses for the
transverse gauge fields. This corresponds to the fact that,
due to 3d gauge and euclidean invariance,
the vacuum polarizations of the zero modes
of the 4d transverse gauge fields in (18) are
not linearly divergent. In the limit of zero external 3--momentum,
$\vec{k}\rightarrow 0$, we have $\Pi^{ab}_{mn}\rightarrow 0$. If this
value is used for the $O(1)$ 3d gauge field contributions to $V_{eff}$
then one obtains contributions to the effective potential proportional
to Tr$\ln(-\vec{\pd}^2+{1\over2}\tg^2\phid\varphi)$, which contains
a cubic term and implies a first order phase transition. However,
this is incorrect.

It was argued in \cite{vj}, that for the most part, the
dominant contributions from the vacuum polarization comes from using the
large momentum limit of $\Pi^{ab}_{mn}$,
\be \Pi^{ab}_{mn}(\vec{k})\rightarrow -\tg^2
   \delta^{ab}(\delta_{mn}-k_m k_n/ \vec{k}^2) {\sqrt{\chi}\over 4\pi},
\ee
not the zero momentum limit. This result is valid for $T\gg
\phid\varphi-\sqrt{\chi}/2\pi\gg \tg^2/(64)^2$ and $\sqrt{\chi}>
\tg^2/(12\pi)$; for technical details the reader is referred to
\cite{vj}. Then the 3d gauge fields give  contributions to $V_{eff}$
proportional to
\be {\rm Tr}\ln \left [-\vec{\pd}^2+{1\over2}\tg^2
       (\phid\varphi-{\sqrt{\chi}\over 2\pi} ) \right ].
             \label{slab} \ee
In a consistent $1/N$ approach, the critical temperature $T_2$
when the origin changes from a local minimum to a local maximum is
again given by the leading order result, $T_2=\sqrt{6}v$ \cite{vj}.
In addition,
to $O(1)$, it is consistent to use the leading order result
(16) for $\sqrt{\chi}$. At $T=T_2$, the combination
$\phid\varphi-\sqrt{\chi}/(2\pi)$ contains no term $\sim \phid\varphi$
for small $\varphi$. Therefore, at $T=T_2$, the effective potential
to $O(1)$ contains no cubic term and the phase transition is
second order \cite{vj}.

At this point we comment on a technical point.
For the limit $\chi\rightarrow 0$ (where are results are not
strictly valid) one can follow \cite{root}.
For the case $g=0$,
Root \cite{root} has shown for the 3d case that the point
$\chi=0$ remains a local minimum at next--to--leading order. This
analysis did not require a computation of $V_{eff}$. Root examined
$dV/d\phi^2$ in the limit of vanishing $\chi$ and showed that the
leading order gap--equation for $\chi$ (in our case eq. (15) )
was sufficient to show that $V_{eff}/d\phi^2=0$ still has a solution
at $\chi=0$ at next--to--leading order. His analysis can be
extended to our gauged case. We do not present a detailed
analysis here but instead refer the reader to \cite{root} and
also ref. \cite{arg} where, following \cite{root}, an analysis of the
small $\chi$ limit in the full 4d high temperature abelian Higgs model
has been performed with the result that $\chi=0$ remains a
point of vanishing $dV_{eff}/d\phi^2$ at next--to--leading order.

This information
is however insufficient to deduce a second order phase transition.
As mentioned in \cite{root}, for sufficiently large $N$ the $O(1)$
corrections cannot overwhelm the leading order conclusion of
a second order phase transition. For $N$ not arbitrarily large
we would like to know the global properties of $V_{eff}$ away
from the point $\chi=0$, and in particular if there is
a point away from the origin at $T=\sqrt{6}v$ where the $O(1)$
corrections can produce a new minimum and possibly even result
in the breakdown of the $1/N$ expansion. In fact, from computing
the effective potential to
$O(1)$ one finds that the ``next--to--leading''
order corrections compete with the ``leading'' corrections when
$g^2/\lambda\sim N$. Thus, the results here are not reliable for in
this case.

The expression (\ref{slab})
appears to characterize, to $O(1)$, the correct behavior in the limit of
vanishing $\chi$ even though it is not strictly valid in this
limit.
Assuming no pathological behaviour occurs
in an exact computation of $V_{eff}$ at next--to--leading order
for  small but nonzero $\sqrt{\chi}$ we
expect that (\ref{slab}) gives a good description all the way down
to $\sqrt{\chi}=0$ and also $\phi^2=0$ at $\sqrt{\chi}=0$.

We now discuss case $ii)$, $g^2/(\lambda N)$ fixed for increasing $N$.
This has the effect of making the pure scalar contributions lower
order in $N$ and, as a result, both the gauge and scalar fields
contribute at
leading order, $O(1)$. To obtain the leading corrections we
make the replacements $\tl=\tl'/N$, $\chi\rightarrow\chi/N$,
$\hc\rightarrow\hc/N$ in (13) and study the limit $\tl'$, $\tg$ fixed
for increasing $N$. For the 3d gauge fields, this has the effect of
making the vacuum polarization contribution in (21) lower order
in $N$. Hence, to $O(1)$, the transverse components of the 4d
gauge fields give contributions to $V_{eff}$ proportional to
Tr$\ln(-\vec{\partial}^2+{1\over 2}\tg^2 \phid\varphi)$. This contains
a cubic term so the phase transition is first order for sufficiently
large $N$.

To conclude,
the high $T$  phase transition in the gauged $SU(2)$ Higgs
model with $N$ complex Higgs doublets is, for sufficiently large $N$,
1st order if $i)$ ${g^2/\lambda}$ grows as $N$ and second order if
$ii)$ ${g^2/\lambda}$ is fixed as $N$ increases. As mentioned, we do
not believe our results can be used to make a reliable statement
about the nature of the phase transition in the electroweak model,
$N=1$, because in this case the number of gauge fields is equal to the
number of real scalars.
In order to make a reliable statement, more subleading corrections than
those that have been computed for either cases $i)$ or $ii)$ must be
determined.
However, our results indicate that if the phase transition is
first order in the exact electroweak model, it is probably more weakly
first order than what the effective potential computed as a sum of ring
diagrams with vacuum polarization effects evaluated only at zero
external 3-momentum would indicate. In this case it is clearly too
weakly first order for generating the BAU [6],[7]. It is therefore
necessary to consider extensions to the standard model in which
the phase transition can be made sufficiently first order in order
to generate the BAU.

The toy models we have studied here and in
refs. \cite{vj},\cite{vjb} demonstrate the care which must be
taken in order to determine the strength of any first order
phase transition. For example, let us add to the tree Lagrangian (1)
an extra scalar particle with the Lagrangian
\be L[s]:= {1\over 2}\partial_\mu s \partial^\mu s
-(m^2+\xi^2\phi^{\dagger}_A
\phi^A)s^2-{\lambda_s\over N} s^4. \ee
Here, $m^2,\xi$ and $\lambda_s$ are field independent.
In the 3d effective theory the one-loop contribution from $s$ to the
effective potential is ${1\over 2}{\rm Tr}\ln (-\vec{\partial}^2+m^2
+\xi^2 \varphi^{\dagger}_A\varphi^A)$. It contains a cubic term
 $\sim\xi^3\varphi^3$ for $m^2$ sufficiently small [8]. However,
since this
extra field for small $m^2$ is very similar to the zero mode of
the longitudinal component of the gauge field in the abelian Higgs
model, it is clear that in the large $N$ limit vacuum polarization
effects will generate a large temperature dependent mass for it and
eliminate the cubic term
for non-negative $m^2$ and $\xi\varphi^2<O(T^2)$.
This conclusion is irrespective of how ${\xi^2/\lambda}$ scales in the
large $N$ limit (with $\lambda_s$ fixed).

Our results suggest
that one should try and find simple extensions of the
$SU(2)$ Higgs model studied here which admit strongly first order phase
transitions in the large $N$ limit. One might hope that, for such models,
corrections to $V_{eff}$ which fall with $N$ cannot significantly reduce
the strength of the first order phase transition, even for small $N$.

\newpage
\begin{center} {\LARGE Figure 1.}\\
     {\Large Gauge loops contributing at O(1).}
\end{center}

\end{document}